\documentclass[a4paper,12pt]{article}
\usepackage{amsmath,amssymb,epsf}
\usepackage{amssymb,amsmath,epsfig,graphicx}
\usepackage{ifthen,theorem,color}

\begin{document}

\begin{center}
{\bf Quantum Energy Eigenvalue Spectra for Very Flat Potentials}\\
Rodney O. Weber\\
School of Mathematics and Applied Statistics\\
University of Wollongong\\
Northfields Avenue, Wollongong, NSW 2522\\
Australia\\
rweber@uow.edu.au\\
\end{center}

{\bf Abstract}

An infinite sequence of potential well functions is considered. A numerical method is 
used for the Schr$\ddot{\text{o}}$dinger equation to obtain the energy eigenvalue  
spectra for a number of these potential well functions. The results for the ground 
state energy are compared to a trial function method. The energy eigenvalue spectra 
are as one would expect although most of them have not been previously reported. 

{\bf 1. Introduction}

The quantum theory of a particle in a potential has been much studied over the last 
hundred years. 
The infinite square well and harmonic potentials are two of the best known examples 
where a complete 
analytic solution is possible and all of the energy levels (eigenvalues) and wave functions 
(eigenfunctions) can be completely determined; e.g. Davies (1990).
Most potential functions do not admit exact analytic solution, but fortunately there are useful
analytic methods that can be used to approximately solve for the eigenvalues and eigenfunctions.
In the current work we will calculate numerical results for the energy eigenvalue spectra  
for a select number of examples of a sequence of potentials, 
beginning with the harmonic potential, each one getting progressively flatter (and for 
comparison purposes we will also list the results for the infinite square well).
The numerical solution is a method from Trefethan (2000) and implemented using $\rm{Octave}^{\rm{TM}}$.  
We will also compare the results for the ground state energies with a trial wavefunction
method used by us previously (Weber, 2018). Our results are as one would expect 
(although most of them have not been previously reported) and are given in tabular form.
An additional useful observation from our work is that the energies are partitioned between two
contributions, similar to a kinetic energy part and a potential energy part and that   
the energies have a predictable, useful, functional dependence.


{\bf 2. Non-dimensionalisation}

We consider the Schr$\ddot{\text{o}}$dinger equation with the same square well sequence 
of potentials as in Weber (2018); namely 
$$
-\frac{\hbar^{2}}{2m}\frac{d^{2}\psi}{dx^{2}}+\mu\left(\frac{x}{a}\right)^{N}\psi=E\psi,  \hskip2cm (1)
$$
where $\mu$ and $a$ are constants and $N$ is a positive natural number, usually assumed 
to be even; however, one can use the absolute value of $\frac{x}{a}$ and raise it to any 
power greater than or equal to 2 and then the Schr$\ddot{\text{o}}$dinger equation (with 
such a potential that employs the absolute value of $\frac{x}{a}$) admits the usual family 
of eigen-wave-function and eigen-energy-value solutions similar to those that are well 
known to anyone who has studied elementary quantum mechanics and the familiar harmonic 
oscillator potential. Our first task is to use non-dimensional variables to re-cast the 
Schr$\ddot{\text{o}}$dinger in as simple a form as possible. To this end one first 
observes that the non-dimensional length $\xi = \frac{x}{a}$ will simplify it to   
$$
-\frac{\hbar^{2}}{2m a^2}\frac{d^{2}\psi}{d\xi^{2}}+\mu \xi^N \psi=E\psi, \hskip2cm (2)
$$
Note that this form of the Schr$\ddot{\text{o}}$dinger equation makes it obvious that 
the physical dimensions of the combination $\frac{\hbar^{2}}{2m a^2}$ is energy, 
the same as the physical dimensions of $\mu$ and $E$; i.e., the physical constants 
in our equation appear to be such that there are two energies that will be involved 
in any expression that we may finally obtain for the energy eigenvalue $E$.

Clearly one can divide by $\frac{\hbar^{2}}{2m a^2}$ in equation (2) and then the 
coefficient in front of $\xi^N$ will be $\mu$ divided by $\frac{\hbar^{2}}{2m a^2}$.
Noticing that there is the second power of $\xi$ involved in the second derivative, 
as well the $N^{\rm{th}}$ power of $\xi$ in the potential suggests that we introduce 
a new constant; let's call it $\sigma$, such that 
$$
\sigma^{N+2} = \frac{\mu}{\frac{\hbar^{2}}{2m a^2}}. \hskip2cm (3)
$$
We can then divide equation by $\sigma^2$ to give
$$
-\frac{1}{\sigma^2}\frac{d^{2}\psi}{d\xi^{2}} + \sigma^N \xi^N \psi = 
\frac{1}{\sigma^2} \frac{E}{\frac{\hbar^{2}}{2m a^2}}\psi. \hskip2cm (4)
$$
Then we rescale with $\xi^* = \sigma \xi$ and arrive at a very tidy expression for
the Schr$\ddot{\text{o}}$dinger equation; namely
$$
-\psi^{''}+\xi^{*N} \psi = \lambda \psi, \hskip2cm (5)
$$
with
$$
\lambda = \frac{1}{\sigma^2} \frac{E}{\frac{\hbar^{2}}{2m a^2}}. \hskip2cm (6)
$$
Equation (5) is to be solved for the eigenvalues $\lambda$ and corresponding eigenfunctions 
$\psi$ for each value of $N$ that is of interest to us. Then the energy eigenvalues for the 
original Schr$\ddot{\text{o}}$dinger equation are obtained by re-arranging equation (6) to 
give
$$
E = \lambda \sigma^2 \frac{\hbar^{2}}{2m a^2}.
$$
And using the definition of $\sigma$ from equation (3) we are able to write the result for
the energy in our final form
$$
E = \lambda (\frac{\hbar^{2}}{2m a^2})^{1-\frac{1}{\beta}} \mu^{\frac{1}{\beta}}. \hskip2cm (7)
$$
Here we have introduced $\beta=\frac{1}{2}(N+2)$ as it allows us to write the result for the
energy in a reasonably compact and neat form. It also facilitates a direct comparison with the
result given in our previous paper (Weber, 2018) where we used a trial wavefunction to 
estimate ground state energies for the Schr$\ddot{\text{o}}$dinger equation (1) for several 
values of $N$ and commented on being able to interpret the result as a partitioning of energy 
into two parts. 

In the present work we numerically solve equation (5) to obtain eigenvalues $\lambda$ for 
several values of $N$. In the case of the lowest eigenvalues (and consequently the ground 
state energies) we will compare the numerical results with the results from the trial 
function approach. Furthermore, we will tabulate the energy levels for several values of $N$ 
in order to clearly see and better understand how the square well sequence influences these 
energy levels. In some respects the results will not be at all surprising: however, it is 
still of interest to see them clearly displayed. Also, we will speculate on how this may be 
relevant to zero point energy calculations, as we intend to follow this up in future work. 

{\bf 3. Numerical Results}

The numerical results are obtained using $\rm{Octave}^{\rm{TM}}$. We have actually used 
$\rm{Matlab}^{\rm{TM}}$ programs from Trefethan (2000) and with minimal adaptation have 
been able to obtain numerical results for the various $N$ values of interest to us in the 
current work. The accuracy can be checked by changing the number of nodes and we have 
been content with reporting the results to four significant figures as we are here 
interested in demonstrating the process and at the same time displaying a table of results 
that illuminate the physical behaviour described by the Schr$\ddot{\text{o}}$dinger equation 
that is of direct interest to us. 

Specific details of the numerical method employed by us in the current work are not our 
original ideas. Rather we have used the programs available in the book by Trefethan (2000) 
(and there are also many more available on his work website as referred to in the book).
The essence of the numerical method is to approximate the second derivative using spectral 
methods and turn the odinary differential equation eigenvalue problem (5) into a matrix 
eigenvalue problem. Background and details for this method for eigenvalue problems 
for ordinary differential equations is discussed and can also be 
found in the references given in Trefethan (2000). 

In table 1 we show the ground state energy eigenvalues obtained by the trial wavefunction 
method as described in Weber (2018) compared to the numerical results. For the harmonic 
oscillator potential ($N=2$) the trial wavefunction happens to be exactly correct and the 
numerical method agrees with the result for the minimum eigenvalue to a high degree of 
accuracy. As $N$ increases (and we have chosen $N=3,4,5,6,7,8$) it is quite obvious that 
the trial wavefunction method consistently returns a result for the ground state eigenvalue 
that is higher than the numerical result and in fact the percentage by which it is higher 
increases as $N$ increases. This is because the trial wavefunctions (generalised Gaussians) 
are poorer approximations to the actual ground state wavefunction as $N$ gets larger. 
From the variational analysis methods for trial wavefunctions, as described in many quantum 
mechanics texts, including Davies (1990), it is possible to prove that the ground state 
energy estimate obtained from using any trial wavefunction will always be greater than 
(or equal to) the actual ground state energy. The results in table 1 are perfectly in 
accordance with this theorem and indeed we are confident that the numerical results for 
the ground state energies are very close to exact.

\begin{table}[h]
\centering
\caption{Ground State Energy Eigenvalues for several values of N} 
\begin{tabular}{c||c|c|c|c|c|c|c} \hline
N & 2 & 3 & 4 & 5 & 6 & 7 & 8 \\ \hline\hline
Trial Wavefunction Method & 1 & 1.053 & 1.157 & 1.288 & 1.434 & 1.592 & 1.758 \\ \hline
Numerical Method & 1.000 & 1.023 & 1.060 & 1.102 & 1.145 & 1.186 & 1.226 \\ \hline
\end{tabular}
\end{table}

In table 2 we show the ground state energies and the first five energies for the excited 
states for $N=2,3,4,5,6,7,8,\infty$. For larger $N$ and higher excited states the numerical 
calculations using our simple method are at first not as accurate, so we have had to 
increase the number of nodes for the spectral methods from 36 to 60 and then through to 80 in 
order to examine the accuracy of the simple numerical method for these higher excited 
states. Other than for $N=2$, 
the eigenvalues are all reported to four significant figures, including the results 
for the infinite square well (infinite $N$) which are obtained from the well known exact 
solution. Importantly, we can clearly see the pattern for the eigenvalue spectra for the 
Schr$\ddot{\text{o}}$dinger equation using our square well sequence of potentials as $N$ 
increases. Namely, larger $N$ means that the spacing between energy eigenvalues gets 
progressively greater (and for comparison we have also tabulated the exact results for the 
energy eigenvalues for the infinite square well which can be obtained by direct solution 
as explained in texts such as Davies, 1990). These results are all exactly as one would 
expect and some of the numbers have been reported before; but to the author's knowledge, 
most of these eigenvalues have not been previously reported.

\begin{table}[!htp]
\centering
\caption{Energy Eigenvalue Spectra Calculated Numerically}
\begin{tabular}{c||c|c|c|c|c|c|c|c} \hline
n::N & 2 & 3 & 4 & 5 & 6 & 7 & 8 & $\infty$ \\ \hline\hline
1 & 1 & 1.023 & 1.060 & 1.102 & 1.145 & 1.186 & 1.226 & 2.467 \\ \hline
2 & 3 & 3.451 & 3.800 & 4.089 & 4.339 & 4.559 & 4.756 & 9.870 \\ \hline
3 & 5 & 6.370 & 7.456 & 8.337 & 9.073 & 9.700 & 10.25 & 22.21 \\ \hline
4 & 7 & 9.522 & 11.64 & 13.43 & 14.94 & 16.23 & 17.34 & 39.48 \\ \hline
5 & 9 & 12.87 & 16.26 & 19.19 & 21.71 & 23.90 & 25.81 & 61.69 \\ \hline
6 & 11 & 16.37 & 21.24 & 25.54 & 29.30 & 32.60 & 35.50 & 88.83 \\ \hline
\end{tabular}
\end{table}

{\bf 4. Conclusion and Further Work}

We have used a numerical method to obtain the energy eigenvalue spectra for a family 
of potentials. The results are as one would expect, although most of them have not 
been previously reported in the literature. 
An additional useful observation from equation (7) is that 
there is a particular combination of the constants in the problems that always
appears in the equation for the eigenvalues and hence the energies; namely
powers of $\hbar^2/2ma^2$ and powers of $\mu$. Checking the dimensions of the result
is easy as $\hbar^2/2ma^2$ has the dimensions of energy, as does $\mu$. And the two powers
always add to one. Hence, it is as though the ground state energy is partitioned between
the two contributions; one kinetic energy and the other potential energy. 
The comparison with a previous trial function method for the ground state energy level 
is also given. Further work on different trial wavefunctions is being considered 
as is the extension of the numerical method to a broader class of potential functions.

Zero point energy calculations, such as in the Casimir effect, begin with the ground state 
energy eigenvalues for the harmonic potential. The results that are then obtained for the 
force between parallel plates have been compared with recent experimental measurements. 
What we wish to stress in this current work, is that the ground state energies for any 
potentials (for example the square well sequence of potential wells considered by us) 
depends on a fundamental length scale in the problem ($a$ in our formulation of the 
Schr$\ddot{\text{o}}$dinger equation) and the potential strength ($\mu$ in our case). 
It is our intention to next repeat some of the zero point energy calculations to see 
how these parameters ($\mu$ and $a$) influence the results. 
 
\newpage

{\bf References}\\
P.C.W. Davies (1984) Quantum Mechanics. Chapman \& Hall (reprinted 1990)\\
N.Trefethan (2000) Spectral Methods in Matlab. SIAM, Philadelphia.\\
R.O.Weber (2018) arXiv:1803.02207\\

{\bf Acknowledgments}\\
The author would like to thank Dr.J.S.Chapman for assistance with LaTeX and R.

{\bf Authors' Contributions}\\
This paper is solely the work of the single author.

{\bf Competing Interests}\\
The author declares no competing interests.

{\bf Data Accessibility}\\
This paper has no data.

{\bf Ethics Statement}\\
There were no ethical considerations required for this work.

{\bf Funding Statement}\\
This work was done without funding.

\end{document}